\documentclass[12pt]{article}
\usepackage{amssymb,amsmath,epsfig,graphicx}
\begin{document}
\title{\bf Noether Symmetry Approach in $f(\mathcal{G},T)$ Gravity }

\author{M. Farasat Shamir \thanks{farasat.shamir@nu.edu.pk}
and Mushtaq Ahmad \thanks{mushtaq.sial@nu.edu.pk}\\\\ National University of Computer and
Emerging Sciences,\\ Lahore Campus, Pakistan.}

\date{}

\maketitle
\begin{abstract}
We explore the recently introduced modified Gauss-Bonnet gravity \cite{sharif.ayesha}, $f(\mathcal{G},T)$  pragmatic with $\mathcal{G}$, the Gauss-Bonnet term, and ${T}$, the trace of the energy-momentum tensor. Noether symmetry approach has been used to develop some cosmologically viable $f(\mathcal{G},T)$ gravity models. The Noether equations of modified gravity are reported for flat FRW universe. Two specific models have been studied to determine the conserved quantities and exact solutions. In particular, the well known deSitter solution is reconstructed for some specific choice of  $f(\mathcal{G},T)$  gravity model.
\end{abstract}


\section{Introduction}

The cosmological aspect of research has proven to be the main resource of modern diverse theories of gravity. No doubt, the onset of General relativity (GR) being a physical theory has been a great success of the last century. However GR alone does not provide us with the sufficient fundamental platform to solve the problems like initial singularity, flatness issues, dark energy and dark matter problems. It also fails when one is interested to express this wide universe as a whole, particularly at the extreme conditions for the ultraviolet scales and for the expression of the quantum structure of space-time. Few years back, the significant outcomes of the researchers have confirmed that the universe is expanding itself \cite{acc}. A gigantic portion of this accelerating universe consists of mysterious substance, known as the dark energy and is believed to be the main cause of the acceleration of this expansion. Recent developments in cosmology have revealed new ideas to introduce the critical and observational innovations for this accelerated expanding universe.

Modified theories of gravity have provided the researchers with different aspects and directions to unveil the hidden realities behind the expansion of the universe. After being motivated and having started with the original theory, a number of  theories have been structured by building intricate Lagrangians. For example, theories of gravity like $f(R)$, $f(R,T)$, $f(\mathcal{G})$ and $f(R,\mathcal{G})$ have been developed by combining curvature scalars, topological invariants and their derivatives. These modified theories address the purpose of solving the complexities related to the quantum gravity and provide with the conditions through which the accelerated expansion of universe is argued. Nojiri and Odintsov \cite{NJO} were the pioneers to present the idea of implicit and explicit coupling of the curvature with the matter in $f(R)$ gravity.
Some reviews \cite{Fel}-\cite{Fel2} were established about the $f(R)$ gravity by different researchers and the consistency of its different cosmological models were also studied \cite{Sta}. The equivalence between metric and Palatini formalisms in $f(R)$ gravity is shown to be achieved using divergence free current \cite{abc}. Harko et al. \cite{HRKO} introduced a gravitational theory by including both the matter and curvature terms and is well known today as the $f(R,T)$ gravity, where $R$ is the scalar curvature and $T$ is the trace of energy-momentum tensor. The evolution of the universe through energy conditions along with the criteria of stability were discussed by Sharif and Zubair \cite{SHZ5}. They also re-established a variety of dark energy models, investigated the thermodynamical aspect and exactly solved the anisotropic universe in $f(R,T)$ gravity \cite{SHZ05}.

In recent years, the researchers have considered a new generalized approach of the Gauss-Bonnet(GB) theory named as $f(\mathcal{G})$ gravity \cite{17}-\cite{18}. The GB invariant is defined as
\begin{equation}
\mathcal{G}=R^{2}-4R_{\alpha\beta}R^{\alpha\beta}+R_{\alpha\beta\gamma\delta}R^{\alpha\beta\gamma\delta},
\end{equation}
where $R_{\alpha\beta\gamma\delta}$ is the Reimann tensor, $R_{\alpha\beta}$ is the Ricci tensor and $R$ is the Ricci scalar.
The interesting feature of the theory is that involvement of GB term may avoid ghost contributions and regularize the gravitational action \cite{Chiba}. Further modification of the theory by involving scalar curvature along with GB invariant has been presented which is named as $f(R,\mathcal{G})$ gravity \cite{NJO6}.
A reasonable amount of work has been published so far in these theories \cite{19}-\cite{Sebastiani}. In a recent paper \cite{sharif.ayesha}, Sharif and Ikram
presented a new modified theory
named as $f(\mathcal{G},T)$ gravity by involving the
trace of the energy-momentum tensor in the function. They investigated
energy conditions for Friedmann–Robertson–Walker (FRW) universe and concluded that the massive test particles followed
non-geodesic lines of geometry due to the presence of extra force. It is expected that this theory may describe the late-time cosmic acceleration for some specific choices of $f(\mathcal{G},T)$ gravity models.

The symmetry methods of approximations have played a pivotal role to work out the exact solutions of  differential equations. These approximations smartly reduce the complexity involved in a system of non-linear equations by finding the unknown parameters of equations. In particular, the Noether symmetries are not just a tool to deal with the solution of the dynamics, but also their existence provides favorable conditions so that we can choose physically and analytically the universe models according to our calculated observations.
Sharif and Waheed \cite{Sharif13} re-scaled the energy of stringy charged black hole solutions using approximate symmetries. Kucukakca \cite{Kucuka16} determined the exact solutions of Bianchi type-$I$ model, using Noether symmetries. Jamil et al. \cite{Jamil17} used the Noether symmetry approach to find out $f(\mathcal{T})$ explicitly for the phantom and quintessence models, where $\mathcal{T}$ is the torsion scalar. Sharif and Shafique \cite{sharif19} discussed  Noether symmetries in a modified scalar-tensor gravity. The exact solutions in $f(R)$ gravity were also explored using Noether symmetries methods for FRW spacetime \cite{Cap21}. Similarly many authors have used Noether symmetries to investigate the cosmology in different contexts \cite{Jamil23}-\cite{abc1}.

In this paper, we are interested to investigate $f(\mathcal{G},T)$ gravity using Noether symmetries. For this purpose, we consider the flat FRW universe model.  The arrangement for the paper is as follows. In section \textbf{2}, we provide the preliminary formalism for $f( \mathcal{G},T)$ gravity. Section \textbf{3} gives the Noether equations of FRW universe model for $f(\mathcal{G},T)$ gravity. Reconstruction of cosmological solutions is presented in section \textbf{4}. Last section provides a brief outlook of the paper.

\section{$f(\mathcal{G},T)$ Gravity with Field Equations}

In 4-dimensions the general action for  $f(\mathcal{G},T)$  gravity
is given by \cite{sharif.ayesha},
 \begin{equation}\label{action}
\mathcal{A}= \frac{1}{2{\kappa}^{2}}\int d^{4}x
\sqrt{-g}[R+f(\mathcal{G},\mathrm{\textit{T}})]+\int
d^{4}x\sqrt{-g}\mathcal{L}_{M},
\end{equation}
where $\mathcal{L}_{M}$ stands for the matter Lagrangian, $R$ is the Ricci Scalar, $\mathcal{G}$ is the Gauss-Bonnet term, $\mathrm{T}$ is the trace of the energy-momentum tensor, $g$ is the metric determinant, and $\kappa$ is the coupling constant.
The energy-momentum tensor $\mathrm{\textit{T}}_{\zeta\eta}$ can be calculated using the following equation:
\begin{equation}\label{emt}
\mathrm{\textit{T}}_{\zeta\eta}=-\frac{2}{\sqrt{-g}}\frac{\delta(\sqrt{-g}\mathcal{L}_{M})}{\delta
g^{\zeta\eta}}.
\end{equation}
However if the distribution of the matter is dependent on the metric tensor $g_{\zeta\eta}$ only then Eq.(\ref{emt}) takes the form
\begin{equation}\label{emt1}
\mathrm{\textit{T}}_{\zeta\eta}=g_{\zeta\eta}\mathcal{L}_{M}-2\frac{\partial\mathcal{L}_{M}}{\partial
g^{\zeta\eta}}.
\end{equation}
Now varying the action Eq.(\ref{action}) with respect to the metric tensor and using Eq.(\ref{emt1}), we obtain the following field equations:
\begin{align}\label{4_eqn}
\begin{split}
G_{\zeta\eta}&=-[2Rg_{\zeta\eta}\nabla^{2}-2R\nabla_{\zeta}\nabla_{\eta}-4g_{\zeta\eta}R^{\mu\nu}\nabla_{\mu}\nabla_{\nu}-
4R_{\zeta\eta}\nabla^{2}+4R^{\mu}_{\zeta}\nabla_{\eta}\nabla_{\mu}\\&
+4R^{\mu}_{\eta}\nabla_{\zeta}\nabla_{\mu}+4R_{\zeta\mu\eta\nu}\nabla^{\mu}\nabla^{\nu}]f_{\mathcal{G}}(\mathcal{G},\mathrm{\textit{T}})+
\frac{1}{2}g_{\zeta\eta}f(\mathcal{G},\mathrm{\textit{T}})-
[\mathrm{\textit{T}}_{\zeta\eta}+\Theta_{\zeta\eta}]f_{\mathrm{\textit{T}}}(\mathcal{G},\mathrm{\textit{T}})\\&
-[2RR_{\zeta\eta}-4R^{\mu}_{\zeta}R_{\mu\eta}-4R_{\zeta\mu\eta\nu}R^{\mu\nu}+2R^{\mu\nu\delta}_{\zeta}R_{\eta\mu\nu\delta}]
f_{\mathcal{G}}(\mathcal{G},\mathrm{T})+\kappa^{2}\mathrm{\textit{T}}_{\zeta\eta},
\end{split}
\end{align}
where ${G}_{\zeta\eta}=R_{\zeta\eta}-\frac{1}{2}g_{\zeta\eta}R$ is the Einstein tensor, $\Box=\nabla_{\zeta}\nabla^{\zeta}=\nabla^{2}$ represents the d'Alembertian operator and $\Theta_{\zeta\eta}= g^{\mu\nu}\frac{\delta
\mathrm{\textit{T}}_{\mu\nu}}{\delta g_{\zeta\eta}}$. Also, $f_{\mathcal{G}}(\mathcal{G},\mathrm{\textit{T}}))$ and $f_{\mathrm{\textit{T}}}(\mathcal{G},\mathrm{\textit{T}}))$ represent the partial derivatives of the function $f( \mathcal{G},T)$ with respect to $\mathcal{G}$ and $\textit{T}$ respectively. It is to be noted that if we take $f(\mathcal{G},\mathrm{\textit{T}})=f(\mathcal{G})$ in Eq.($\ref{4_eqn}$), then the resulting equation represents the field equations of $f(\mathcal{G})$ gravity. Moreover, by putting $f(\mathcal{G}, \mathrm{\textit{T}})=0$, one can recover the usual Einstein field equations.
The trace of Eq.($\ref{4_eqn}$) yields
\begin{align}\label{5_eqn}
\begin{split}
&R+\kappa^{2}\mathrm{\textit{T}}-(\mathrm{\textit{T}}+\Theta)f_{\mathrm{\textit{T}}}(\mathcal{G},\mathrm{\textit{T}})+2f(\mathcal{G},\mathrm{\textit{T}})+2\mathcal{G}f_{\mathcal{G}}(\mathcal{G},\mathrm{\textit{T}})-2R\nabla^{2}f_{\mathcal{G}}(\mathcal{G},\mathrm{\textit{T}})\\&
+4R^{\zeta\eta}\nabla_{\zeta}\nabla_{\eta}f_{\mathcal{G}}(\mathcal{G},\mathrm{\textit{T}})=0.
\end{split}
\end{align}
This is an important equation as it can be used to find the corresponding $f(\mathcal{G},\mathrm{\textit{T}})$ models.
The covariant divergence of Eq.(\ref{4_eqn}) is given as
\begin{equation}\label{div}
\nabla^{\zeta}T_{\zeta\eta}=\frac{f_{\mathrm{\textit{T}}}(\mathcal{G},\mathrm{\textit{T}})}
{\kappa^{2}-f_{\mathrm{\textit{T}}}(\mathcal{G},\mathrm{\textit{T}})}\bigg[(\mathrm{\textit{T}}_{\zeta\eta}+\Theta_{\zeta\eta})
\nabla^{\zeta}(\text{ln}f_{\mathrm{\textit{T}}}(\mathcal{G},\mathrm{\textit{T}}))+
\nabla^{\zeta}\Theta_{\zeta\eta}-
\frac{g_{\zeta\eta}}{2}\nabla^{\zeta}T\bigg],
\end{equation}
The theory might be plagued by divergences, e.g. at astrophysical scales, due to the presence of higher-order derivatives of the stress-energy tensor that are naturally present in the field equations. This seems to be an issue with higher order derivatives theories that includes higher order terms of stress-energy tensor. Modification of Einstein's theory by adding auxiliary fields does not compromise the weak equivalence
principle and admits a covariant Lagrangian formulation \cite{auxiliary}. However one can put some constraints to Eq.(\ref{div}) to obtain standard conservation equation for stress-energy tensor \cite{sharif.ayesha}.

In this paper, we restrict ourselves to flat Friedmann-Robertson-Walker (FRW) spacetime,
\begin{equation}\label{13_eqn}
ds^{2}=dt^{2}-a^{2}(t)(dx^{2}+dy^{2}+dz^{2}).
\end{equation}
The standard matter energy-momentum tensor is defined as
\begin{equation}\label{6}
T_{\mu\nu}=(\rho + p)u_\mu u_\nu-pg_{\mu\nu},
\end{equation}
satisfying the EoS
\begin{equation}\label{eos}
p=w\rho,
\end{equation}
where $u_\mu=\sqrt{g_{00}}(1,0,0,0)$ is the four-velocity in
co-moving coordinates and $\rho$ and $p$ denote energy density
and pressure of the fluid respectively.
The scheme of our paper can be mainly divided into two points as follows:
\begin{itemize}
  \item Equations (\ref{4_eqn}) are highly non-linear differential equations and its not an easy task to solve them analytically.
  It is mentioned here that fiat FRW modified field equations are fourth-order differential equations involving unknowns like $f(\mathcal{G}, \mathrm{\textit{T}})$.  Here we are interested to find the Noether symmetries of $f(\mathcal{G},T)$ gravity with fiat FRW background.
  \item The advantage of exact solutions in modified gravity have gained much importance, particularly in the study of phase transitions and recent phenomenon of accelerated expansion of universe. The viable cosmological models can be found using Noether symmetries and hence some physically important solutions can be reconstructed.
\end{itemize}

\section{Noether Symmetries and $f(\mathcal{G},T)$ Gravity}

Noether symmetries have become an important tool to solve the system of non-linear equations. We apply Noether symmetry approach to investigate the $f( \mathcal{G},T)$ gravity. The existence of this approach confirms the uniqueness of the vector field in the tangent space. In this situation, the vector field acts like symmetry generator which produces further the conserved quantities. The expression for the vector field and its  first prolongation are given respectively as \cite{Olver}
\begin{equation}\label{204_eqn}
\mathrm{\textit{W}}=\zeta(t,u^{j})\frac{\partial}{\partial
t}+\xi^{i}(t,u^{j})\frac{\partial}{\partial u^{j}},
\end{equation}
\begin{equation}\label{24_eqn}
\mathrm{\textit{W}}^{[1]}=\mathrm{\textit{W}}+(\xi^{i}_{~,t}+\xi^{i}_{~,j}~\dot{u}^{j}-\zeta_{~,t}~\dot{u}^{i}-\xi_{~,j}~
\dot{u}^{j}\dot{u}^{i})\frac{\partial}{\partial\dot{u}^{j}}.
\end{equation}
where $\zeta$ and $\xi$ are the coefficients of the generator, $u^i$ provides the $n$ number of positions, and the dot gives the derivative with respect to time $t$. The vector field $\textit{W}$ produces Noether gauge symmetry provided the condition
\begin{equation}\label{25_eqn}
\mathrm{\textit{W}}^{[1]}\mathcal{L}+(D\zeta)\mathcal{L}=DG(t,u^{i})
\end{equation}
is preserved, where $G(t,u^i)$ denotes gauge term and $D$ is an operator defined as
\begin{equation*}\label{26_eqn}
D=\frac{\partial}{\partial
t}+\dot{u}^{i}\frac{\partial}{\partial u^{i}}.
\end{equation*}
The Euler-Lagrange equations are given by
\begin{equation}\label{29_eqn}
\frac{\partial \mathcal{L}}{\partial
u^{i}}-\frac{d}{dt}\Bigg(\frac{\partial \mathcal{L}}{\partial
\dot{u}^{i}}\Bigg)=0,~~~~~
\end{equation}
Contraction of Eq.(\ref{29_eqn}) with some unknown function $\phi^{i}\equiv\phi^{i}(u^{j})$ yields
\begin{equation}
\phi^{i}\Big(\frac{\partial \mathcal{L}}{\partial
u^{i}}-\frac{d}{dt}\Big(\frac{\partial \mathcal{L}}{\partial
\dot{u}^{i}}\Big)\Big)=0.\label{12}
\end{equation}
It is easy to verify that
\begin{equation}
\frac{d}{dt}\Big(\phi^{i}\frac{\partial \mathcal{L}}{\partial
\dot{u}^{i}}\Big)-\Big(\frac{d}{dt}\phi^{i}\Big)\frac{\partial
\mathcal{L}}{\partial \dot{u}^{i}}
=\phi^{i}\frac{d}{dt}\Big(\frac{\partial \mathcal{L}}{\partial
\dot{u}^{i}}\Big).\label{13}
\end{equation}
Putting this value in Eq.(\ref{12}) provides us with
\begin{equation}\label{abc}
L_\textit{W}\mathcal{L}
=\phi^{i}\frac{\partial \mathcal{L}}{\partial
u^{i}}+\Big(\frac{d}{dt}\phi^{i}\Big)\frac{\partial
\mathcal{L}}{\partial
\dot{u}^{i}}=\frac{d}{dt}\Big(\phi^{i}\frac{\partial
\mathcal{L}}{\partial \dot{u}^{i}}\Big),
\end{equation}
where $L$ stands for the Lie derivative along the vector field.
The Noether symmetries would exist only if the Lie derivative of the Lagrangian becomes zero, i.e.
the condition
\begin{equation*}\label{28_eqn}
L_{\mathrm{\textit{W}}}\mathcal{L}=0.
\end{equation*}
Since the Lagrangian $\mathcal{L}$ is invariant along the vector field
$\textit{W}$, consequently the definition for Noether current turns out to be \cite{Scap2 Note}
\begin{equation}
j^{t}=\Big(\phi^{i}\frac{\partial \mathcal{L}}{\partial
\dot{u}^{i}}\Big),\label{16}
\end{equation}
and for the Noether current to be a conserved quantity, we must have
\begin{equation}\label{17}
j^{t}_{,t}=0.
\end{equation}
Now we write again the action (\ref{action}) for the case of perfect fluid as
\begin{equation}\label{32_eqn}
\mathcal{A}=\int dt\sqrt{-g}[R+f(\mathcal{G},\mathrm{\textit{T}})-\mu_{1}(\mathcal{G}-\mathcal{\bar{G}})-\mu_{2}(\mathrm{\textit{T}}-\bar{T})+\mathcal{L}_{M}].
\end{equation}
Here $\textit{T}$ is the trace of energy-momentum tensor, $\sqrt{-g}=a^{3}(t),\bar{\mathcal{G}}$ and $\bar{\mathrm{\textit{T}}}$ stand for dynamical constraints, while $\mu_{1}$ and $\mu_{2}$ are the multipliers, worked out as
\begin{equation*}\label{33_eqn}
\begin{split}
\mu_{1}=f_{\mathcal{G}}(\mathcal{G},\mathrm{\textit{T}}),~~
\mu_{2}=f_{\mathrm{\textit{T}}}(\mathcal{G},\mathrm{\textit{T}}),~~
\bar{\mathrm{\textit{T}}}=\rho(a)-3p(a).
\end{split}
\end{equation*}
Since there does not exist any unique definition of matter Lagrangian, so we consider $\mathcal{L}_{M}=-p(a)$ and after an integration by parts, the point-like Lagrangian takes the following form
\begin{eqnarray}\nonumber
&\mathcal{L}(a,R,\mathcal{G},\mathrm{\textit{T}},\dot{a},\dot{R},\dot{\mathcal{G}},\mathrm{\textit{T}})=
a^{3}[R+f-\mathcal{G}f_{\mathcal{G}}
-f_{\mathrm{\textit{T}}}\{\mathrm{\textit{T}}-(\rho(a)-3p(a))\}-p(a)]\\\label{34_eqn}&
-8\dot{a}^{3}\dot{\mathcal{G}}f_{\mathcal{GG}}-
8\dot{a}^{3}\dot{\mathrm{\textit{T}}}f_{\mathcal{G}\mathrm{\textit{T}}},
\end{eqnarray}
where $f\equiv f(\mathcal{G})$, $f_{\mathcal{G}}\equiv f_{\mathcal{G}}(\mathcal{G},\mathrm{\textit{T}})$ etc.
Using Eq.(\ref{34_eqn}), the Euler-Lagrangian equations turn out to be
\begin{equation}\label{35_eqn}
\begin{split}
&3a^{2}[R+f-\mathcal{G}f_{\mathcal{G}}-
f_{\mathrm{\textit{T}}}\{\mathrm{T}-(\rho(a)-3p(a))\}-p(a)]\\&
+a^{3}[-f_{\mathrm{\textit{T}}}\{-3p_{,a}(a)+\rho_{,a}(a)\}-p_{,a}(a)]-
[\dot{\mathcal{G}}^{2}f_{\mathcal{G}\mathcal{G}\mathcal{G}}+2\dot{\mathcal{G}}\dot{\mathrm{\textit{T}}}f_{\mathcal{G}\mathcal{G}\mathrm{\textit{T}}}\\&
\ddot{\mathcal{G}}f_{\mathcal{G}\mathcal{G}}+\dot{\mathrm{\textit{T}}}^{2}f_{\mathcal{G}\mathcal{G}\mathrm{\textit{T}}}+\ddot{\mathrm{\textit{T}}}
f_{\mathcal{G}\mathrm{\textit{T}}}]=\frac{2\ddot{a}}{\dot{a}}\bigg(\dot{\mathcal{G}}f_{\mathcal{G}\mathcal{G}}+\dot{\mathrm{\textit{T}}}
f_{\mathcal{G}\mathrm{\textit{T}}}\bigg),
\end{split}
\end{equation}
\begin{equation}\label{36_eqn}
\bigg(\frac{\dot{a}}{a}\bigg)^{2}\bigg(\frac{\ddot{a}}{a}\bigg)=-\frac{1}{24f_{\mathcal{GG}}}\bigg
[-\mathcal{G}f_{\mathcal{GG}}-\mathrm{\textit{T}}f_{\mathcal{G}}-
f_{\mathcal{G}\mathrm{\textit{T}}}\{\mathrm{\textit{T}}-(\rho(a)-3p(a))\}\bigg],
\end{equation}
\begin{equation}\label{37_eqn}
\bigg(\frac{\dot{a}}{a}\bigg)^{2}\bigg(\frac{\ddot{a}}{a}\bigg)=-\frac{1}{24f_{\mathcal{G}\mathrm{\textit{T}}}}
\bigg[f_{\mathrm{\textit{T}}}-
\mathcal{G}f_{\mathcal{G}\mathrm{\textit{T}}}-f_{\mathrm{\textit{TT}}}
\{\rho(a)-3p(a)\}-f_{\mathrm{\textit{T}}}\bigg].
\end{equation}
Also the corresponding vector field takes the form
\begin{equation}\label{37_eqn}
\mathrm{\textit{W}}=\alpha\frac{\partial}{\partial a}+\beta\frac{\partial}{\partial R}+\gamma\frac{\partial}{\partial \mathcal{G}}+\delta\frac{\partial}{\partial \mathrm{\textit{T}}}+\dot{\alpha}\frac{\partial}{\partial \dot{a}}+\dot{\beta}\frac{\partial}{\partial \dot{R}}+\dot{\gamma}\frac{\partial}{\partial \dot{\mathcal{G}}}+\dot{\delta}\frac{\partial}{\partial \dot{\mathrm{\textit{T}}}},
\end{equation}
where $\alpha,~\beta,~\gamma$ and $\delta$ are functions of $a,~R,~\mathcal{G}$ and $\mathrm{\textit{T}}$.
Now using Lagrangian $(\ref{34_eqn})$ and Noether equation $(\ref{25_eqn})$, an over-determined system of partial differential equations (PDE's) is obtained: 
\begin{equation}\label{14_eqn}
-8\frac{\partial \gamma}{\partial
R}f_{\mathcal{G}\mathcal{G}}-8\frac{\partial \delta}{\partial
R}f_{\mathcal{G}\mathrm{\textit{T}}}=0,
\end{equation}
\begin{equation}\label{15_eqn}
-8\gamma f_{\mathcal{G}\mathcal{G}\mathcal{G}}-8\delta
f_{\mathcal{G}\mathcal{G}\mathrm{\textit{T}}}-24\frac{\partial
\alpha}{\partial a}f_{\mathcal{G}\mathcal{G}}-8\frac{\partial
\gamma}{\partial
\mathcal{G}}f_{\mathcal{G}\mathcal{G}}-8\frac{\partial
\delta}{\partial \mathcal{G}}f_{\mathcal{G}\mathrm{\textit{T}}}=0,
\end{equation}
\begin{equation}\label{16_eqn}
-8\gamma
f_{\mathcal{G}\mathcal{G}\mathrm{\textit{T}}}-8f_{\mathcal{G}\mathrm{\textit{T}}\mathrm{\textit{T}}}-24\frac{\partial
\alpha}{\partial
a}f_{\mathcal{G}\mathrm{\textit{T}}}-8\frac{\partial
\gamma}{\partial
\mathrm{\textit{T}}}f_{\mathcal{G}\mathcal{G}}-8\frac{\partial
\delta}{\partial
\mathrm{\textit{T}}}f_{\mathcal{G}\mathrm{\textit{T}}}=0,
\end{equation}
\begin{equation}\label{17_eqn}
-8\frac{\partial \gamma}{\partial
a}f_{\mathcal{G}\mathcal{G}}-8\frac{\partial \delta}{\partial
a}f_{\mathcal{G}\mathrm{\textit{T}}}=0,
\end{equation}
\begin{equation}\label{18_eqn}
-24\frac{\partial \alpha}{\partial
\mathcal{G}}f_{\mathcal{G}\mathrm{\textit{T}}}-24\frac{\partial
\alpha}{\partial \mathrm{\textit{T}}}f_{\mathcal{G}\mathcal{G}}=0,
\end{equation}
\begin{equation}\label{19_eqn}
-24\frac{\partial \alpha}{\partial
R}f_{\mathcal{G}\mathrm{\textit{T}}}=0,
\end{equation}
\begin{equation}\label{20_eqn}
-24\frac{\partial \alpha}{\partial R}f_{\mathcal{G}\mathcal{G}}=0,
\end{equation}
\begin{equation}\label{21_eqn}
-24\frac{\partial \alpha}{\partial
\mathcal{G}}f_{\mathcal{G}\mathcal{G}}=0,
\end{equation}
\begin{equation}\label{22_eqn}
-24\frac{\partial \alpha}{\partial
\mathrm{\textit{T}}}f_{\mathcal{G}\mathrm{\textit{T}}}=0,
\end{equation}
\begin{align}\label{23_eqn}
\begin{split}
&3a^{2}\alpha
[R+f-\mathcal{G}f_{\mathcal{G}}-f_{\mathrm{\textit{T}}}
\{\mathrm{\textit{T}}-(\rho(a)-3p(a))\}-p(a)]\\&
-a^{3}\alpha[f_{\mathrm{\textit{T}}}\{-3p_{,a}(a)+\rho_{,a}(a)\}-p_{,a}(a)]+a^{3}\beta+\gamma
a^{3}[-\mathcal{G}f_{\mathcal{G}\mathcal{G}}
-f_{\mathcal{G}\mathrm{\textit{T}}}\times\\&\{\mathrm{\textit{T}}-(\rho(a)-3p(a))\}]+
\delta a^{3}[-\mathcal{G}f_{\mathcal{G}\mathrm{\textit{T}}}-\mathrm{\textit{T}}f_{\mathrm{\textit{T}}
\mathrm{\textit{T}}}+f_{\mathrm{\textit{T}}\mathrm{\textit{T}}}(\rho(a)-3p(a))]=0.
\end{split}
\end{align}
It is mentioned here that for the sake of simplicity, we have considered the gauge term zero.
Using Eq.(\ref{17}), conservation equation for Noether charge takes the form
\begin{eqnarray}\label{18}
\frac{d}{dt}\Bigg[\alpha\Big\{\frac{\partial}{\partial
\dot{a}}\big(8\dot{a}^{3}\dot{\mathcal{G}}f_{\mathcal{GG}}+
8\dot{a}^{3}\dot{\mathrm{T}}f_{\mathcal{G}\mathrm{T}}\big)\Big\}+
\gamma\frac{\partial}{\partial
\dot{\mathcal{G}}}(8\dot{a}^{3}\dot{\mathcal{G}}f_{\mathcal{GG}})-\delta\frac{\partial}{\partial
\dot{\mathrm{T}}}(\dot{a}^{3}\dot{\mathrm{T}}f_{\mathcal{G}\mathrm{T}})\Bigg]=0.
\end{eqnarray}
Now we solve the system of PDE's (\ref{14_eqn}-\ref{23_eqn}) for different cases.\\\\
$\mathbf{Case(i)}$:\\\\
Let us assume that $f_{\mathcal{G}\mathcal{G}}=0$. Manipulating the Eqs.(\ref{14_eqn}-\ref{23_eqn}), we obtain $\alpha=0$, $\beta=0$, $\gamma=0$, and $\delta=0$, hence providing us with a trivial solution. Moreover, the conservation equation (\ref{18}) is also satisfied in this case. Thus, we have to consider $f_{\mathcal{G}\mathcal{G}}\neq0$ to obtain a non-trivial solution.
\\\\
$\mathbf{Case(ii)}$:\\\\
Now let us assume that $f_{\mathcal{G}\mathcal{\textit{T}}}=0$ and $f_{\mathcal{G}\mathcal{G}}\neq0$.
Therefore, the cosmological model takes the form $f(\mathcal{G},\textit{T})=a_{0}\mathcal{G}^{2}+b_{0}\textit{T}^{2}$, where $a_{0}$ and $b_{0}$ are the arbitrary constants. Here the manipulation of Noether equations provides $\alpha=0=\gamma$, $\delta=c_{1}$ and
$\beta=c_{1}\textit{T}+c_{2}$. Thus the symmetry generator becomes
\begin{equation}
\textit{W}=(c_{1}\textit{T}+c_{2})\frac{\partial}{\partial R}+c_{1}\frac{\partial}{\partial\textit{T}}.
\end{equation}
Now the conservation equation (\ref{18}) referring to the Noether current gives
\begin{equation}\label{con}
\dot{a}^{3}f_{\mathcal{G}\mathrm{T}}=c_3,
\end{equation}
where $c_3$ is an integration constant. It is mentioned here that this case satisfies the conservation equation (\ref{con}) when we choose the integration constant equal to zero. The corresponding Lagrangian becomes
\begin{align}\label{44_eqn}
\begin{split}
\mathcal{L}&=a^{3}[R-a_{0}\mathcal{G}^{2}-b_{0}\mathrm{\textit{T}}^{2}+2b_{0}\mathrm{\textit{T}}\{\rho(a)-3p(a)\}-p(a)]-16\dot{a}^{3}a_{0}\dot{\mathcal{G}}.
\end{split}
\end{align}
The corresponding Euler-Lagrangian equations are calculated as
\begin{align}\label{42_eqn}
\begin{split}
&a^{2}(R-a_{0}\mathcal{G}^{2}-b_{0}\mathrm{\textit{T}}^{2}+2b_{0}\{\rho(a)-3p(a)\}\mathrm{\textit{T}}-p)+32a_{0}\dot{a}\ddot{a}\dot{\mathcal{G}}+16a_{0}\dot{a}^{2}\ddot{\mathcal{G}}=0,\\
&a^{3}(R-2a_{0}\mathcal{G})+48a_{0}\dot{a}^{2}\ddot{a}=0,~~~~-2a^{3}b_{0}(T-\{\rho(a)-3p(a)\})=0.
\end{split}
\end{align}
Using Eq.(\ref{42_eqn}), it follows that
\begin{equation}
a(t)=c_{4}\sqrt{2t+c_{5}}, ~~~~ \mathcal{G}=\frac{-24}{(2t+c_{5})^{4}}.
\end{equation}
where $c_{4}$, and $c_{5}$ are the constants of integration.

\section{Reconstruction of Cosmological Solutions}

In this section, we provide the cosmological solutions by considering $f(\mathcal{G},\textit{T})=\mathcal{G}^{k}\textit{T}^{1-k}$, where $k$ is an arbitrary real number. We reconstruct the solutions for the case ${k}=2$.
The Lagrangian takes the shape
\begin{equation}\label{41_eqn}
\mathcal{L}=a^{3}\bigg[R-\frac{\mathcal{G}^{2}\{\rho(a)-3p(a)\}}{\mathrm{\textit{T}}^2}-p(a)\bigg]-\frac{16\dot{a}^{3}}{\mathrm{\textit{T}}}\bigg(\dot{\mathcal{G}}-\frac{\dot{\mathrm{\textit{T}}}}{\mathrm{\textit{T}}}\bigg).
\end{equation}
Euler-Lagrangian equations here take the form
\begin{eqnarray}\nonumber
&3a^{2}\bigg[R-\frac{\mathcal{G}^{2}\{\rho(a)-3p(a)\}}{\mathrm{\textit{T}}^2}-p(a)\bigg]-a^{3}\bigg[
\frac{\{\rho(a)-3p(a)\}_{,a}\mathcal{G}^{2}}{\mathrm{\textit{T}}^2}+p_{,a}(a)\bigg]\\\label{a}&
+\frac{48(\dot{a}^{2}\ddot{\mathcal{G}}-2\dot{a}\ddot{a}\dot{\mathcal{G}})}{\mathrm{\textit{T}}}-\frac{48\dot{\mathrm{\textit{T}}}
\dot{a}^{2}\dot{\mathcal{G}}}{\mathrm{\textit{T}}^{2}}-\frac{48(\dot{a}^{2}\ddot{{\mathrm{\textit{T}}}}+2\dot{\textit{T}}\dot{a}\ddot{a})}{\mathrm{\textit{T}}^{2}}+\frac{96{{\dot{{\mathrm{\textit{T}}}}}^2\dot{a}}^2}{\mathrm{\textit{T}}^{3}}
=0,\\\label{b}&
\mathcal{G}\{\rho(a)-3p(a)\}-24\mathrm{\textit{T}}\dot{a}^{2}\ddot{a}+8\dot{a}^{3}\dot{\mathrm{\textit{T}}}=0,\\\label{49_eqn}&
{a}^{3}\mathcal{G}^{2}\{\rho(a)-3p(a)\}-8\mathrm{\textit{T}}\dot{a}^{3}
\dot{\mathcal{G}}+24\mathrm{\textit{T}}\dot{a}^{2}\ddot{a}=0.
\end{eqnarray}
Using Eq.(\ref{49_eqn}) and by putting the corresponding values for $\mathcal{G}$ and $\mathrm{\textit{T}}$, we get
\begin{equation}\label{41_eqn}
40a{\ddot{a}}^2{\dot{a}}^4+8a{\dot{a}}^5{\dddot{a}}-24{\dot{a}}^6{\ddot{a}}-a^4{\ddot{a}}{\dot{a}}^2=0.
\end{equation}
This equation admits a solution of exponential form
\begin{equation}
a=e^{mt},
\end{equation}
where $m$ is an arbitrary constant. It is mentioned here that exponential solution is satisfied with the constraint equation
\begin{equation}
24m^8-m^4=0
\end{equation}
yielding the real solutions
\begin{equation}
m=0, ~~~m\approx 0.451801,~~~m\approx -0.451801.
\end{equation}
Also, using Eq.(\ref{b}) the value of trace of energy momentum tensor turns out to be
\begin{equation}
T=c_5e^{\frac{3}{4}(4-a^{3})},
\end{equation}
where $c_5$ is constant of integration.
Using Eq.(\ref{a}) and Eq.(\ref{b}), we obtain
\begin{equation}\label{d}
p(a)_{,a}+\frac{3}{a}p(a)=l(a),
\end{equation}
where
\begin{equation}
l(a)=\frac{c_5e^{\frac{3}{2}(a^{3}-4)}}{a^{3}}\bigg[-3a^2{\mathcal{G}}^2\textit{T}+a^3\mathcal{G}^2{\textit{T}}_a+
96\dot{a}\textit{T}(\ddot{a}\dot{\mathcal{G}}-\dot{a})+48\dot{a}^2(\dot{\mathcal{G}}\dot{\textit{T}}-\ddot{\mathcal{G}}\textit{T}
+\ddot{\textit{T}})-\frac{96(\textit{T}\dot{a})^2}{\textit{T}}\bigg].
\end{equation}
It is evident that Eq.(\ref{d}) is a non-homogeneous linear differential equation and one can solve it to find the pressure and
consequently the energy density of the universe.
Thus in this case the solution metric takes the form
\begin{equation}\label{62a}
ds^{2}=d{t}^{2}-e^{2mt}[dx^{2}+d{y}^{2}+d{z}^{2}].
\end{equation}
This corresponds to the well-known deSitter spacetime in GR.
Here we have constructed a physical cosmological solution with a particular $f(\mathcal{G},\textit{T})$ gravity model.
Similarly, more solutions with some other cosmological models can be reconstructed.

\section{Outlook}

In this paper, we have discussed in detail about the Noether symmetries of the flat FRW universe model in $f(\mathcal{G},\mathrm{T})$ gravity. Noether symmetries are not just a tool to deal with the solution of the dynamics, but also their existence provides favorable conditions so that we can choose physically and analytically the universe models according to our calculated observations. Lagrangian multipliers perform a big part to shape the Lagrangian into its canonical form and so as to reduce the dynamics to determine the exact solutions. We have worked out the Lagrangian for FRW universe model in $f(\mathcal{G},\mathrm{T})$ theory.  The existence of Noether charges is considered important in the literature and equation for conservation of charge  plays an important role to investigate the Noether symmetries. The conservation equation for Noether charge has been developed. The exact solutions of Noether equations have been discussed for two cases of $f(\mathcal{G},\mathrm{T})$ gravity models. The first case when $f_{\mathcal{G}\mathcal{G}}=0$ yields trivial symmettries while we obtain non-trivial symmetries for the second case when $f_{\mathcal{G}\mathrm{\textit{T}}}=0$ and $f_{\mathcal{G}\mathcal{G}}\neq0$. Thus we have also worked out the corresponding $f(\mathcal{G},\mathrm{T})$ gravity model and the solution metric. It is concluded that the second case provides $f(\mathcal{G},\textit{T})=a_{0}\mathcal{G}^{2}+b_{0}\textit{T}^{2}$ gravity model, where $a_{0}$ and $b_{0}$ are arbitrary constants. Furthermore, solutions in both cases satisfy the conservation equation for Noether charge. 

We have also reconstructed an important cosmological solution by considering $f(\mathcal{G},\textit{T})=\mathcal{G}^{k}\textit{T}^{1-k}$, where $k$ is an arbitrary real number. This model yields the well-known deSitter solution already available in GR.
It is mentioned here that many other cosmologically physical solutions may be reconstructed for some other choice of $f(\mathcal{G},\textit{T})$ gravity models.
We have discussed the exact solutions with only three cases. Many other solutions can be explored by assuming some other forms of $f(\mathcal{G},\mathrm{T})$.
\\\\
\textbf{Acknowledgement}\\\\ Many thanks to the anonymous reviewer
for valuable comments and suggestions to improve the paper. 
This work was supported by National University
of Computer and Emerging Sciences (NUCES).

\end{document}